\documentclass[apj]{emulateapj}
\usepackage{amsmath}
\usepackage{graphicx}
\usepackage{epstopdf}
\usepackage{hyperref}
\usepackage[usenames,dvipsnames]{color}  
\usepackage{natbib}
\newcommand{\kms}{\ensuremath{\,\rm{km}\,\rm{s}^{-1}}}
\newcommand{\Msun}{\ensuremath{\,M_\odot}}
\newcommand{\Lsun}{\ensuremath{\,L_\odot}}
\newcommand{\kpc}{\ensuremath{\,\rm{kpc}}}
\newcommand{\fbin}{\ensuremath{f_{\rm{bin}}}}

\begin{document}
\title{ The incidence of stellar mergers and mass gainers among massive stars  }

\author{
S.E.~de Mink\altaffilmark{1,2*},   H. Sana\altaffilmark{3},  N.~Langer\altaffilmark{4} ,   R.G.~Izzard\altaffilmark{4} \& F.R.N. Schneider\altaffilmark{4} 
} 

\altaffiltext{1}{Observatories of the Carnegie Institution for Science, 813 Santa Barbara St, Pasadena, CA 91101, USA}
\altaffiltext{2}{Cahill Center for Astrophysics, California Institute of Technology, Pasadena, CA 91125, USA}
\altaffiltext{3}{Space Telescope Science Institute, Baltimore, MD 21218, USA }
\altaffiltext{4}{Argelander Institut f\"ur Astronomie der Universit\"at Bonn, Germany}
\altaffiltext{*}{Einstein Fellow.}

\begin{abstract}
Because the majority of massive stars are born as members of close binary systems, populations of massive main-sequence stars contain stellar mergers and products of binary mass transfer. We simulate populations of massive stars accounting for all major binary evolution effects based on the most recent binary parameter statistics and extensively evaluate the effect of model uncertainties.  

Assuming constant star formation, we find that $8^{+9}_{-4}\,\%$  of a sample of early type stars to be the product of a merger resulting from a close binary system. In total we find that  $30^{+10}_{-15}\,\%$ of massive main-sequence stars are the product of binary interaction.  

We show that the commonly adapted approach to minimize the effects of binaries on an observed sample by excluding systems detected as binaries through radial velocity campaigns can be counterproductive. Systems with significant radial velocity variations are mostly pre-interaction systems.  Excluding them substantially enhances the relative incidence of mergers and binary products in the non radial velocity variable sample. 

This poses a challenge for testing single stellar evolutionary models. It also raises the question of whether certain peculiar classes of stars, such as magnetic O-stars, are the result of binary interaction and it emphasizes the need to further study the effect of binarity on the diagnostics that are used to derive the fundamental properties (star-formation history, initial mass function, mass to light ratio) of stellar populations nearby and at high redshift.

\end{abstract}


\section{Introduction}

Young massive stars are predominantly found in close binary systems  \citep[e.g.][]{Mason+2009, Sana+2011, Chini+2012}. This implies that the majority of massive stars interacts with a companion before ending their lives as supernovae \citep{Sana+2012}.   Such interaction has drastic consequences for the further evolution and final fate of both stars \citep[e.g.][]{Podsiadlowski+1992, Wellstein+1999, Eldridge+2008} and gives rise to a variety of exotic phenomena including blue stragglers, X-ray binaries, mili-second pulsars and gamma-ray bursts.   

The strong preference for close systems, with orbital periods of less than a few days, implies that a third of the systems interacts while both stars are still on the main sequence \citep{Sana+2012}.    Binary evolutionary models predict that a large fraction of these systems evolve into a contact configuration \citep[e.g][]{Pols1994a, Wellstein+2001, Nelson+2001, de-Mink+2007}.  The evolution of massive contact binaries is uncertain, but it is anticipated that the stars are driven deeper into contact due to the continuing expansion of the stars and shrinkage of the orbit due to angular momentum loss.  In particular, the loss of mass with high specific angular momentum through the outer Lagrangian point, possible torques resulting from circum-binary material and dissipative processes occurring in the common envelope are believed to eventually result in a merger of the two stars \citep[e.g.][]{Podsiadlowski+1992, Wellstein+2001}.  

Recently, the merger of a close binary system, although at lower mass, was caught in the act as the transient V1309 Sco \citep{Tylenda+2011}.  It has also been suggested that the light echo of V838 Mon resulted from a massive merger \citep{Munari+2002, Tylenda+2006}, but this case remains more controversial.  

For massive stars the fraction of stars that will merge as a result of contact in a close binary is predicted to be as high as 20-30\% according to the most recent binary statistics \citep{Sana+2012}.  In addition, mergers may be triggered by dynamical interactions between three bodies systems \citep[e.g.][]{Kozai1962, Perets+2009, Hamers+2013} or as a result of (multiple) collisions in dense star clusters \citep{Portegies-Zwart+2004,Glebbeek+2009} or near the galactic center \citep{Antonini+2010, Antonini+2011}. 

The chance of witnessing the merger event of two massive stars is small, because of the scarcity of massive stars.  The rate of such events in our galaxy is expected to be about 20--30\% of the Galactic supernova rate, about one every two hundred years \citep[e.g.][]{Langer2012}.  In contrast, the products of such mergers may be rather common, in particular for mergers between two main sequence stars. These objects are expected to be rejuvenated as fresh fresh is mixed into the central burning regions \citep[e.g.][]{Glebbeek+2013}.  These merger products are expected to be among the brightest stars in young clusters forming a massive analogue of blue stragglers \citep[e.g.][Schneider et al. 2013a, subm.]{Mermilliod1982, Chen+2009a, Lu+2010}. 

For the population of early-type stars in a typical galaxy, which is not characterized by a single burst of star formation, mergers and mass gainers cannot easily be identified.  However, if such stars are abundant, they can in principle affect various diagnostics that are used to derive the fundamental properties such as the star-formation history and initial mass function \citep[e.g.][]{van-Bever+1998, Eldridge2012,Zhang+2012, Li+2012}.  Such properties are generally derived using population synthesis models such as GALAXEV \citep{Bruzual+2003}, STARBURST99 \citep{Leitherer+2010} and FSPS \citep{Conroy+2009}, in which all stars are in principle assumed to evolve in isolation.   Estimating the incidence of binary products is therefore ultimately motivated by the need to improve our understanding of the validity of the properties derived for stellar populations nearby as well as those at high redshift. 

A more direct motivation is the need to test state-of-the-art stellar evolution models, which contain
uncertain effects of convection, rotation and magnetic fields \citep[e.g.][]{Brott+2011, Ekstrom+2012, Potter+2012}.  Lacking prescriptions from first principles makes calibration against observed populations indispensable.  These models generally assume the stars to evolve in isolation.  However, interaction with a binary companion can lead to drastic changes in the observable properties.  To evaluate the validity of tests and calibrations of the models against observed samples, it is necessary to estimate the contamination of such samples with stars that  are the products of binary interaction. 

In this work we take a first step towards quantifying the implications of the newly derived binary fraction and distribution of orbital properties of massive binary stars \citep{Sana+2012}.  For this purpose we employ a rapid synthetic binary evolution code that has been updated to adequately describe the main relevant processes in Sect.~\ref{Smethod}. We simulate a population of young massive stars in a typical galaxy assuming  continuous star formation to compute the incidence of mergers and other products of binary evolution.  In particular, we examine which binary products can be detected as binary systems through radial velocity variations in Sect.~\ref{Sbias}.  We assess the significance of our results by varying uncertain input distributions and the adopted treatment of uncertain physical processes in Sect.~\ref{Suncertainties}. In Sect.~\ref{Sobs} we discuss the presence of binary products in observed samples and how they may be recognized and we conclude in Sect.~\ref{Scon}.

\section{Method} \label{Smethod}

To estimate the effects of binary interaction on a population of early-type stars  we adopt the binary fraction, and  distribution of orbital periods and mass ratios based on observations of O-stars in nearby ($\lesssim 3\kpc$) young (about 2\,Myr) star clusters, which have been subject to an intense spectroscopic monitoring campaign \citep{Sana+2012}. 
 Although the sample is of modest size, including 71 systems containing at least one O-star, it exceeds any previous samples in completeness providing orbital solutions for over 85\% of the detected binary systems.  This allowed the derivation of the distribution of binary parameters corrected for incompleteness and biases, resulting in the following distribution of initial orbital periods, $p$, \[
 f_p (\log p)  \sim ( \log p )^\pi
 \]
for $\log p {\rm {(days)}} \in [0.15, 3.5]$, where $\pi = -0.55\pm 0.2$  and a distribution of mass ratios $q$ defined as the mass of the less massive star over the mass of the more massive star,   
\[
f_q (q)  \sim q^\kappa
\]
for $q \in [0.1, 1]$ where $\kappa = -0.1 \pm 0.6$.  The binary fraction, i.e. the number of binary system with periods and mass ratios in the range specified with respect to the total number of single and binary systems, is $f_{\rm bin} = 0.69 \pm 0.09$.   The distribution of primary masses  is consistent with a mass function, \[
f_m (m) \sim m^{-\alpha}
\]
with $\alpha=2.35$ \citep{Salpeter1955}. Given the young ages of the clusters in this sample, we consider these in our simulations as initial conditions at the onset of hydrogen burning. 

In our standard simulation we adopt ${f_{\rm bin} = 0.7}$, ${\pi = -0.5}$, ${\kappa = 0}$ and ${\alpha = 2.35}$.  We we consider variations on these parameters (see Tab.~\ref{tab}) that generously include the confidence interval quoted by \citet{Sana+2012}.  The remaining fraction  $1-f_{\rm bin}$ is included as single stars, even though these may in reality have a nearby low mass companion or a companion in a wide orbit.  
 
To model the effect of stellar evolution and binary interaction we employ a synthetic binary evolution code that is described in detail in  \citet{de-Mink+2013-paper} and references therein (hereafter Paper I).  This code was originally developed by \citet{Hurley+2000, Hurley+2002} and \citet{Izzard+2004, Izzard+2006, Izzard+2009} based on stellar models by \citet{Pols+1998}. It has been updated and extended to include various processes relevant for this study (Paper I).   A brief summary of the main aspects is given here. 

We account for mass and angular momentum loss through winds  \citep{Nieuwenhuijzen+1990, Vink+2001},  effects of rotation on the stellar winds \citep{Maeder+2000} and deformation due to rotation in the Roche approximation (Paper I). We adopt a metallicity of $Z=0.02$. We account for interaction through tides \citep{Zahn1977, Hurley+2002} and Roche lobe overflow \citep{Hurley+2002} assuming circular orbits.  In our standard simulation, we limit the accretion rate by the thermal rate of the accreting star \citep{Tout+1997,Hurley+2002} and we assume that the remainder of the mass is lost from the system taking away the specific orbital angular momentum of the accreting star.  Because the efficiency of mass transfer, $\beta$, defined as the fraction of the mass transferred from the donor star to the companion star that is actually accreted by the companion, 
and the specific angular momentum, $\gamma$, of material lost from the system are highly uncertain, we consider the extreme cases of conservative mass transfer ($\beta = 1$) and the highly non-conservative case where a star can no longer accrete after it has been spun up to its Keplerian limit (c.f.\ Paper I). This latter case is indicated as $\beta = \beta_{\rm K}$ and is equal to the assumptions made in \citet{Petrovic+2005} and \citet{de-Mink+2009}.  For the angular momentum loss we consider the extreme case that all mass is lost through the outer Lagrangian point (indicated as $\gamma = \gamma_{L}$) and the extreme limit where the specific angular momentum of the mass lost from the system is negligible ($\gamma = 0$).

We assume that binary systems come in contact when  $M_{\rm acc}/M_{\rm don} < q_{\rm crit}$ \citep[e.g.][]{Kippenhahn+1977}, where $M_{\rm acc}$ and $M_{\rm don}$ refer to the mass of the accreting star and the donor star.  For systems with a main sequence donor star we adopt $q_{\rm crit, MS} = 0.65$, based on a calibration against a grid of detailed binary evolutionary models \citep{de-Mink+2007}.   Given the considerable uncertainties concerning the formation of contact (c.f. Sect 5.2.1 in Paper I) we explore the extreme cases of $q_{\rm crit, MS} \in \{0.25,0.75\}$.  For systems with a Hertzsprung-gap donor star we adopt $q_{\rm crit, HG} = 0.4$ and explore the effect of the extreme assumptions $q_{\rm crit, HG} \in \{0.0,q_{\rm W01}\}$, where $q_{\rm W01} = 1.0$, for  $p > 30$\,d and 0.65 for $p \le 30$\,d to mimic the detailed models by \citet{Wellstein+2001}.  We follow \citet{Hurley+2002} for the treatment of evolved donors.

We assume that contact binaries merge. During the coalescence we assume in our standard simulations that a fraction $\mu_{\rm loss}=0.1$ of the system mass is lost during the merger  taking away the excess angular momentum.  When investigating the uncertainties we consider the case that mass loss can be neglected,  $\mu_{\rm loss}=0$, and $\mu_{\rm loss}=0.25$.  The merger product is assumed to settle to its thermal equilibrium structure while rotating near its Keplerian rotational velocity. It is assumed that the core of the most evolved star, which has the lowest entropy, sinks to the center of the merged star.   In our standard simulation only a small fraction $\mu_{\rm mix}=0.1$ of the hydrogen-rich envelope is mixed into the new convective core \citep[e.g.][]{Gaburov+2008, Glebbeek+2013}.  We also consider the extreme case that the merger product is completely mixed $\mu_{\rm mix} = 1$, as is assumed in the original version of the code, and the case of no additional mixing $\mu_{\rm mix} = 0$.  For accreting stars we account for rejuvenation by assuming that the star adapts it structure to its new mass, mixing in fresh hydrogen as the convective core expands (Paper I and references therein).   

In our simulations of a population of early-type stars we select stars that are undergoing central hydrogen burning for which the combined brightness of the main-sequence components exceeds $10^4\Lsun$ (see Paper I).  We chose this approach instead of estimating the spectral types directly from the effective temperatures, because our predictions of the luminosity and evolutionary phase are more reliable than effective temperatures.  The adopted limit corresponds in our models to the brightness of a single main-sequence star with initial mass $\gtrsim 8{\text{--}}12\Msun$, where the range reflects the fact that stars become brighter as they evolve over the main sequence.   This roughly corresponds to stars of spectral types early B and O.  We consider continuous star formation, which is a good approximation for a large system with multiple bursts of star formation, such as the complete early-type star population in a typical Galaxy.  The case of different star formation histories (starbursts) computed with the same code and very similar assumptions are discussed in \citet[][subm.]{Schneider+2013}.

\begin{figure}[t]\center
   \includegraphics[ width=0.5\textwidth]{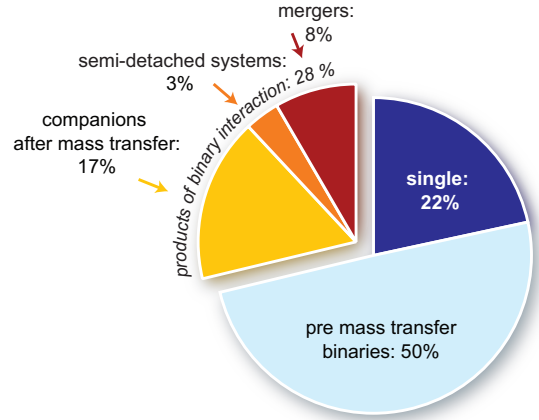}
  \caption{ The incidence of stellar mergers and (post) mass transfer systems in our standard simulation of a population of massive main sequence stars assuming continuous star formation and an initial binary fraction of 70\%. Percentages are expressed in terms of the number of systems containing at least one main sequence star. See Sec.~\ref{Sinc} for a discussion and Table~\ref{tab} for the impact of model uncertainties on these predictions.
    \label {f0} \label{pie_all}} 
\end{figure}

\begin{figure*}[t!]\center
  \includegraphics[width=\textwidth]{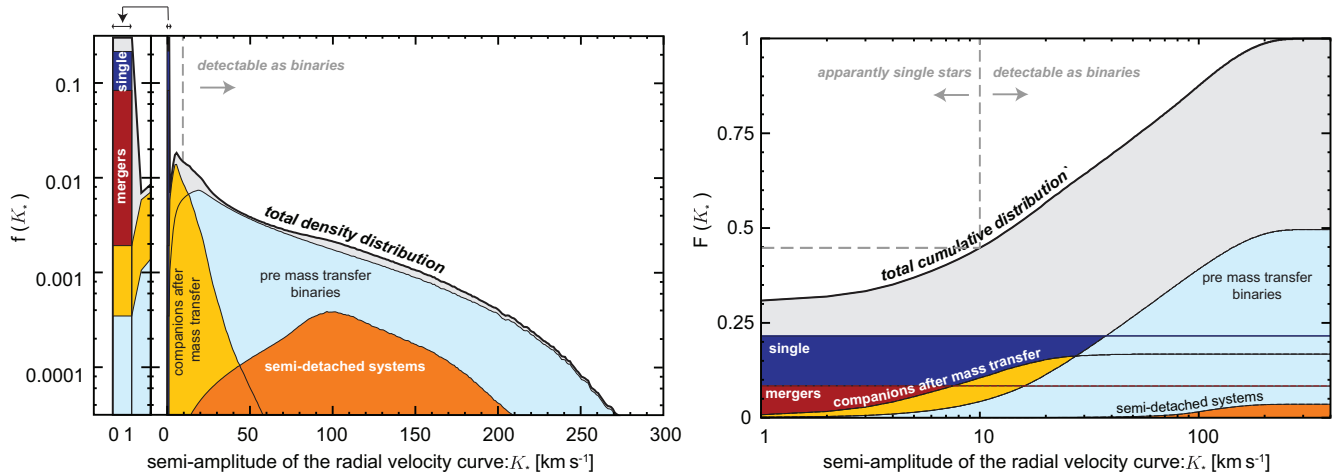}
  \caption{ The density and cumulative distribution of the semi-amplitudes, i.e.\ the maximum line-of-sight velocities due to the orbital motion, for a population of massive early-type stars including a fraction of binaries characteristic for the Milky Way.   The distribution is normalized after adding single stars and mergers.  Note that we adopted logarithmic-linear and linear-logarithmic scales to enhance different features. 
   \label {dists}} 
\end{figure*}

\section{The incidence of binary products in a population of massive early-type stars} \label{Sinc}
The properties of our simulated stellar population are summarized in Fig.~\ref{f0}, which indicates the relative contribution of single stars, binary stars and products of binary interaction. The percentages are expressed in terms of the number of systems, either single or binary, that contain at least one main sequence star and for which the combined brightness of the main-sequence components exceeds $10^4\Lsun$. A binary system containing two main-sequence stars is counted only once.  

The simulations started with 70\% binaries and 30\% single stars at birth.  For the case of continuous star formation we find that the contribution of pre interaction binaries is reduced from 70\% at birth to 50\% in the current population as a result of stellar evolution and binary interaction.  About a fifth consists of stars that we refer to as single stars, even though some may have a companion in a very wide orbit, or very low mass companion, which are not accounted for in our simulation.   

More than a quarter of the systems have been severely affected by interaction with their companion.  This group mainly consists of mergers (8\%) and stars that have previously gained mass from a (former) companion (17\%). The fraction of systems that are in a semi-detached configuration, i.e. that are currently undergoing mass transfer through Roche-lobe overflow, is small (3\%).  This is because the mass transfer phase typically lasts for a few thermal timescales at most, which is short compared to the stellar lifetime.   An exception are close systems that experience case A mass transfer \citep{Nelson+2001}, i.e.\ mass transfer from a main-sequence donor, which can last for up to about a third of the main sequence lifetime \citep[e.g. Fig.~3 in ][]{de-Mink+2007}.    Practically all semi-detached systems, accounting for a few percent of the population, are undergoing slow Case A mass transfer.  These systems form a subset of the observed algol type systems. 

The large fraction of mergers and mass gainers can be understood as the combined effect of different mechanisms.  First,  their production is favored by the large close binary fraction at birth.   Second, binary products have gained mass through accretion or coalescence, resulting in an increase in brightness.   Stars that were initially not massive and luminous enough to be included in our brightness limited sample can become bright enough after mass accretion.  In other words, the binary products in our sample come from a wider range of initial masses than the single stars and stars that have not yet interacted. In addition, these binary products originate from lower mass systems, which are favored by the slope of the initial mass function.   Third, when a main sequence star accretes mass it typically adapts its internal structure leading to an increase of the size of the convective core. As a result the hydrogen rich layers above the original core are mixed to the central burning regions providing fresh fuel, which effectively rejuvenates the star \citep{Kippenhahn+1977, Podsiadlowski+1992, Braun+1995,Dray+2007,Claeys+2011}.  The prolonged lifetime increases the fraction of stars that is expected to be observed after mass transfer.

\subsection {The counter-productive effect of selecting against binaries. \label{Sbias}}

The contamination of a sample of early-type stars with binary products poses a challenge for their usefulness to test stellar models.  A commonly adopted approach to try to reduce the effects of binaries on an observed sample is to exclude every object that is a known binary.  Here, we demonstrate that removing detected binaries from a sample is counter-productive \citep[cf.][]{de-Mink+2011}.

Spectroscopic binaries are detected through variations in the radial velocity resulting from the orbital motion. In a single-lined circular spectroscopic binary, the maximum radial velocity variation, $\Delta v$, that can be measured if the orbital phase is well sampled near quadrature, is equal to  $2 K_*$, where $K_*$ is the semi-amplitude of the radial velocity curve for the brightest star,  and  $i$ is the inclination angle.  In a typical early-type system with a primary mass $M_*$,  a mass ratio $q$ and an orbital period $p$, 
\begin{eqnarray}
K_*  \approx  80\kms &&   \bigg(\frac{\sin i}{ \pi/4 } \bigg)
  \bigg( \frac{M_*}{20\Msun}  \bigg)^{\alpha_m} \bigg( \frac{q}{0.5}  \bigg)^{\alpha_q}   \bigg( \frac{p}{10\, {\rm d}}  \bigg)^{\alpha_p}   \nonumber
 \end{eqnarray} 
 where  $\alpha_m  = 1/3$ and $\alpha_p  = -1/3$ and the mass ratio dependence is approximated as a power law with exponent $\alpha_q \approx 0.86$ for $q \in [0.05,1]$.  

Obtaining accurate radial velocity measurements for early-type stars is challenging because there are relatively few suitable lines.   Moreover, the lines are typically broad as a result of rotation, and as a result of macro and micro turbulence.  A typical accuracy that can be reached when measuring radial velocity variations is 1--10\kms~depending on the signal to noise ratio, the spectral type and the rotation rate.  Pulsations and stellar winds can further induce apparent variations of low amplitude.  Therefore, radial velocity variations of $\Delta v \ge \Delta v_{\lim} \equiv 20\kms$  (or $K_* \ge 10\kms$) is typically considered as an unambiguous sign of orbital motion due to the presence of a companion \citep[e.g.][]{Sana+2012a}. 

In Fig.~\ref{dists} we show the probability density function $f$ and the cumulative distribution $F$ of semi-amplitudes of the binaries in our simulated population assuming random inclination angles.   The distributions are normalized $f$ after adding single stars and mergers, for which we set  $K_* = 0\kms$.  The detection limit is indicated as a vertical dashed line. As can been seen in the right panel, about 45\% of the objects do not show any radial velocity variations that are large enough to be un ambiguously detected as caused by a companion star. This group consists of stars that appear to be single.  

The remaining 55\% show radial velocity variations larger than the detection limit.  In principle these are detectable as binaries. This requires multiple spectra to be taken that cover enough different phases of the orbit such that the full radial velocity curve can be reconstructed.   For semi-amplitudes larger than the detection limit, the distribution is dominated by the primary stars of binary systems that have not yet interacted by mass transfer.  Their semi-amplitudes extend beyond 250 \kms, although semi-amplitudes of around 20\kms, approaching the detection limit, are most common. 
Systems that are currently undergoing Roche-lobe overflow typically have semi-amplitudes around 100\kms, well above the detection limit.  This group is dominated by mass transfer systems in which the Roche lobe filling star is a main-sequence star. These systems have compact orbits leading to large radial velocity variations. 

\begin{figure*}[t!]\center
   \includegraphics[ width=\textwidth]{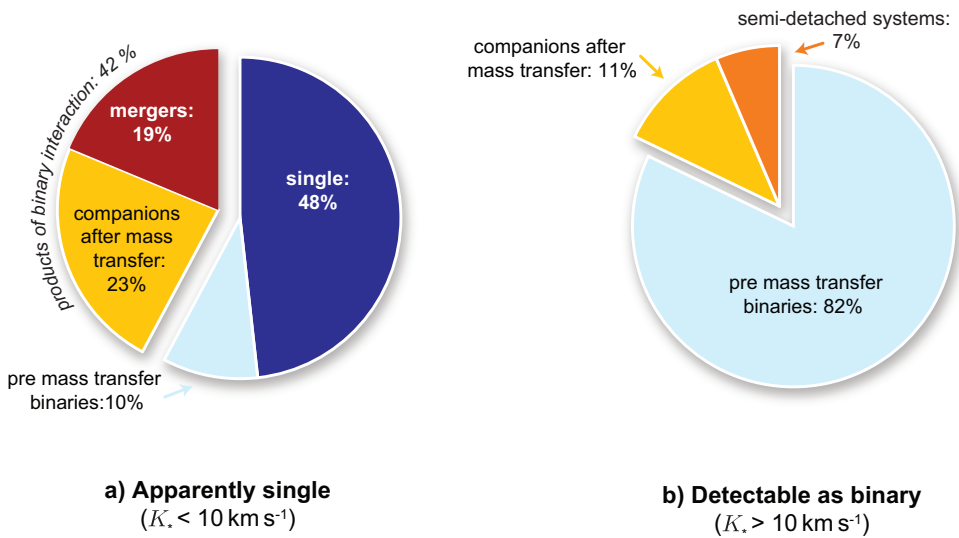}
  \caption{ Selecting against stars with evidence for a companion in order to reduce the contamination of binaries in an observed sample can be counter productive.  A sample of apparently single stars ($K_*  < 10 \kms$) contains a large fraction of mergers and other binary products, cf Fig.~\ref{f0} (left pie chart).  The systems that are removed are preferentially binary systems that have not yet experienced interaction (right pie chart). 
    \label {piecharts}\label{pie_single}} 
\end{figure*}

The post interaction systems indicated as ``companions after Roche-lobe overflow" consist of a main-sequence star accompanied by a helium star, a neutron star, black hole or white dwarf.   These systems typically have small semi-amplitudes that are below the unambiguous detection limit.  A significant fraction will be unbound as a result of the birth kick of the neutron star during the supernova explosion of the primary.  Because the dependence of the birth kick distribution on the pre-explosion properties is highly uncertain, we assume a zero kick velocity in all cases, such that the quoted number of post-interaction systems with detectable radial velocity variations provides an upper limit.

When removing systems from a sample,  that are detected as radial velocity variables, one preferably removes binary systems from the sample that have not yet interacted.  The products of binary interaction, which typically do not show measurable radial velocity variations, are left in the sample.  Fig.~\ref{piecharts} summarizes the consequences for an idealized observing campaign, in which sufficient observing time is allocated to acquire full phase coverage of each system. In this case each binary with a semi-amplitudes $K_* > \frac{1}{2}\Delta {v}_{\lim} \equiv 10\kms$ would be identified.    

The left pie chart shows the population of stars with no significant radial velocity variations, i.e stars that appear to be single. About half of this sample are indeed stars were born as single stars (48\%). A small fraction (10\%) corresponds to pre-interaction binary systems displaying radial velocity variations that are too small to be detected.  This group consists of the widest binaries and binaries with orbits that are aligned nearly face.  These two groups together, accounting for 58\% of the sample, are stars that have lived their lives so far without experiencing any significant interaction with a companion. 

The remaining 42\% consists of stars whose evolutionary history deviates strongly from that of an isolated star.  These consist of mergers, (19\%) and stars that have gained mass through Roche lobe overflow in the past (27\%).  In reality systems with radial velocity variations approaching the detection limit are hard to detect without extensive monitoring.   The pie chart on the righthand side shows that those binaries that in principle can be detected through radial velocity variations. This sample consists preferentially of pre-interaction systems. 

We conclude that excluding detected binaries from a sample to reduce the contamination of binary products is counter-productive. The binary products are typically not detected and will therefore represent a larger fraction of the sample than single stars.    Those objects that are identifiable as binaries are dominated by stars that have not yet interacted.   Apart from the closest systems in which tides play a role these systems have lived their lives effectively as single stars.

\subsection{Effect of model uncertainties \label{Suncertainties}}

In Table~\ref{tab} we summarize the effect of varying the initial conditions and the main uncertain physical assumptions on the incidence of binary products  and mergers in particular. 

Concerning the adopted initial distributions of binary parameters, it is the power law exponent in the distribution of orbital periods, which is the main cause for uncertainty. We find that the fraction of stars that are binary products varies from 19 to 40\% when considering an initial distribution that is flat in $\log p$ (i.e. $\pi = 0$ or ``Opik's law'') and a distribution that strongly favors short period systems ($\pi = -1$).  Similarly, the incidence of merger products (accounting for 8\% in our standard simulation) varies by about a factor of two up or down (4 - 17\%) when varying the slope of the initial period distribution.  

Varying the mass ratio distribution from a distribution that strongly favors systems with unequal masses ($\kappa = -1$) to a distribution that favors systems with equal masses ($\kappa = 1$)  increases the fraction of binary products from 22 to 33\%.  This results from the fact that binaries with comparable masses produce more massive binary products that are luminous enough to meet our brightness limit.   The fraction of mergers varies only from 10 to 7\%. This is because of two effects that partially compensate each other.  Systems with unequal masses are more likely to come in contact and merge.  However, if systems with comparable masses merge they produce brighter objects.  We refer to the discussion section in Paper I for further details.  

The effect of changing the initial mass function more modest. We find a variation of 27 to 32\% when adopting flatter ($\alpha = 1.65$) or a steeper ($\alpha = 3.05$) steeper slope for the initial stellar mass function, respectively.   When changing the adopted initial binary fraction down to $f_{\rm bin} = 0.3$ or up to $f_{\rm bin} = 1.0$ we find changes in the contamination by binary products between 15\% and 37\%.   When changing the adopted brightness limit from $10^4$ to $10^5\Lsun$ we find a increase of the incidence of binary products (to 34\%) and mergers (to 12\%). 

Investigating the effects of the uncertain physical assumptions requires recomputing the grid of models.  To speed up the computations we recomputed them at lower resolution. In Table~\ref{tab} we provide the results for variations of the physical assumptions as well as a reference simulation at the same resolution adopting our standard assumptions. 

The dominant uncertainty affecting the predicted number of binary products is the accretion efficiency. This still remains one of the major uncertainties in binary evolutionary models.   Attempts to constrain the efficiency using large samples of binary systems with accurately determined parameters remained inconclusive \citep[e.g.][and references therein]{de-Mink+2007}.  

If we assume that all the mass lost by the primary star through Roche-lobe overflow is accreted by the companion we find that the fraction of binary products increases to 34\%, because of the larger masses and thus luminosities of the binary products.  Adopting highly non-conservative mass transfer by assuming that stars do not accrete after they have been spun up to their Keplerian rotation rate, which is typically after they have gained a few percent of their mass \citep[][]{Packet1981}, reduces the incidence of binary products to 18\%.   Changing the adopted specific angular momentum of material that is lost from the system during interaction affects the number of mergers and whether the systems display detectable radial velocity variations after interaction.  However, we find that this effect is minor compared to the other uncertain assumptions.

The number of mergers is mainly affected by the amount of mixing in the merger product. When the merger product is assumed to be fully mixed, the remaining lifetime is enhanced as a result of the fresh hydrogen that has become available in the central burning regions.  The fraction of mergers is also sensitive to the adopted critical mass ratio for contact in systems where the donor star is on the main-sequence.  These two uncertainties are responsible for variations of 7-12\%.  We find that our predictions are not very sensitive to the adopted amount of mass loss during the merger event, nor to the adopted critical mass ratio for contact in systems where the donor star is a post main-sequence object.

\begin{table*}[t!]
\caption{Impact of uncertainties in the initial conditions and physical assumptions on the incidence of mergers and the total number of binary products among a population of massive early-type stars.  Symbols are explained in Sect.~\ref{Smethod}. The underlined values are the largest deviation in each category.  \label{tab}}
\begin{center}
\begin{tabular}{llcc|cc}
\tableline\tableline
 \multicolumn{2}{c}{} & standard &  extreme  & \multicolumn{1}{c}{all binary products$^*$ } &  \multicolumn{1}{c}{mergers only} \\
 \multicolumn{2}{c}{} &values &  values& \multicolumn{1}{c}{ (\%)} &  \multicolumn{1}{c}{(\%)} \\

\tableline
\multicolumn{4}{l|}{\bf Initial conditions} & \\
&\multicolumn{3}{l|}{Standard simulation for reference}&  28.7 & ~8.4 \\ 
&- $\quad$primary mass distribution & $\alpha =2.35$  &  $ 1.65,3.05$   &  26.8 -- 32.1 &  ~7.8 -- ~9.4  \\ 
&- $\quad$mass ratio distribution & $\kappa = 0 $ &    $-1,1$               &   22.3 -- 33.0 &  ~9.9 -- ~7.1  \\  
&- $\quad$orbital period distribution & $\pi = -0.5$&    $0,-1  $              &  19.4 -- {\underline{39.9}}  &   ~\underline{3.5} -- \underline{16.6}  \\
&- $\quad$binary fraction & $\fbin = 0.7$&   $ 0.3,1.0$        & \underline{14.7} -- 36.5&   ~4.3 -- 10.7  \\
&- $\quad$metallicity &  $Z = 0.2$& $ 0.004, 0.03$ & 26.5 -- 36.8 & ~4.5 -- ~9.0   \\
&- $\quad$brightness limit &  $L_{\rm lim} (\Lsun) = 10^4$& $ 10^5$ & 34.2 & 12.2     \\


\multicolumn{4}{l|}{\bf Main physical assumptions}&\\
& \multicolumn{3}{l|}{Standard simulation for reference}&  27.3 & ~8.0  \\
&- $\quad$accretion efficiency & $\beta = \beta_ {\rm th}$& $  \beta_ {\rm K}, 1 $  &     \underline {17.7} -- \underline {34.3}    &     ~7.4 -- ~7.2  \\

&- $\quad$angular momentum loss & $\gamma = \gamma_{acc}$& $ 0, \gamma_{\rm L}$ & 27.4 -- 24.5 & ~8.1 -- ~8.4  \\
 &- $\quad$contact during main sequence   & $q_{\rm crit, MS} = 0.65$&    $0.25, 0.8$ & 28.1 -- 26.4    &  ~\underline{6.6} -- ~9.1   \\
 &- $\quad$contact during Hertzsprung gap & $q_{\rm crit, HG} = 0.4$&$0, q_{W01}$ &  21.7 -- 23.0 & ~8.5 -- ~8.4  \\
 &- $\quad$treatment mergers: mass loss & $\mu_{\rm loss} = 0.1$&  $0, 0.25$  & 27.1 -- 27.3  & ~7.8 -- ~8.0  \\
 &- $\quad$treatment mergers: mixing &  $\mu_{\rm mix} = 0.1$ &  $0,1$& 26.6 -- 30.6 & ~7.1 -- \underline{12.3} \\
\hline
\end{tabular}
 \end{center}
$^*$All binary products, i.e. mergers, systems undergoing Roche-lobe overflow and companions after Roche-lobe overflow  
\end{table*}

\section{Binary products in observed samples}\label{Sobs}

The number of binary products present in observed stellar samples depends on the selection effects and the details of the underlying star-formation history.   In populations younger than 2\,Myr one might expect the overall contamination by binary products to be small, since only the most massive stars evolve fast enough to expand and interact. However, depending on the selection criteria, the contamination by binary products may very significant.  Even after only 1-2\,Myr, the brightest star of a well populated star cluster is expected to be the product of binary evolution \citep{Schneider+2013}.  Therefore, in a sample consisting of the brightest stars of young star clusters, the contamination by mergers and other binary products may be significantly larger than predicted by our simulations for continuous star formation.  

As discussed in Section~\ref{Sbias}, the inclusion or selection against known binaries can significantly change the relative number of binary products.  Modeling this effect is very challenging since the selection against binaries is often not done in a systematic way.  For example, in a typical sample serendipitously discovered binaries described in the literature are excluded.  Given the limited size of most samples these unsystematic selection effects can not be ignored. Because of these sample dependent difficulties, a quantitative comparison of our predictions with observations beyond the scope of this paper. However, we will make several general remarks.  

An example of a large homogenous survey of massive stars is the VLT-FLAMES Tarantula Survey (VFTS) of massive stars  \citep{Evans+2011}. This ESO large program is a multi-epoch spectroscopic survey of 800 randomly selected early-type stars, among which 360 O stars, in the Tarantula Nebula or 30 Doradus region of the Large Magellanic Cloud \citep{Walborn1984}.  The sample contains a mix of stellar populations with different ages including one that is at least ~20 Myr old. 

The fraction of O stars in this survey that are detected to be spectroscopic binaries $0.35 \pm 0.03$ \citep{Sana+2012a}. Even though this multi-epoch spectroscopic survey was designed to systematically search for binaries, it is far from the idealized case demonstrated in Fig.~\ref{piecharts}b.   After carefully modeling the specific biases fro this survey, \citet{Sana+2012a} derive an intrinsic binary fraction of $0.51 \pm 0.04$.   Based on our simulations we expect the contribution of post interaction systems among the detected binaries to to be very small. One may therefore tentatively compare this to the fraction of pre-interaction systems in our simulation for continuous star formation, e.g. Fig.~\ref{pie_all}, where we find a remarkably similar fraction of 50\%.  

The intrinsic binary fraction derived for the Tarantula Survey is lower than that derived for young galactic clusters, which is $0.7\pm0.1$ \citep[][]{Sana+2012}.   The difference is small and might be purely the result of stochastic effects.  However, our simulations show that this difference comes out naturally as a result of stellar evolution and binary interaction. The O stars in the young galactic survey have not had time to interact. Their binary fractions is expected to closely resemble conditions at birth, i.e. the onset of hydrogen burning. Instead, the Tarantula Survey contains early-type stars of a wide range of ages. The effects of stellar evolution and binary interactions have modified the intrinsic binary fraction. 

A different large survey is the first VLT-FLAMES survey of massive stars \citep{Evans+2005}, containing stars in our galaxy, the small and the large Magellanic cloud.  This sample seems to be reasonably well characterized by continuous star formation.  \citep{Brott+2011a} used the early B-type stars in this survey to calibrate overshooting and the efficiency of rotationally induced mixing in stellar models for isolated single stars.  Detailed modeling of the selection effects is required to estimate the contamination of binary products in this sample. However, based on our simulations one may expect the contaminated fraction to be in the order of almost a third of the sample. This is similar to the total fraction of stars with enhanced nitrogen abundances  \citep[box 1 and 3 in Fig.~10 in ][which account for 32.8\% of the sample]{Brott+2011a}.  This questions whether the observed nitrogen abundances are the result of rotational mixing processes operating in isolated rotating stars, or whether they are the result of binary interaction processes \citep{Hunter+2008a}.

Finally, a highly tentative comparison, which should be taken with great care, is the fraction of binary products or mergers with the observed fraction of peculiar stars.  For example, the fraction of massive stars that are observed to be magnetic stars , i.e.\ about 10\%, \citep{Donati+2009, Hubrig+2011, Hubrig+2013, Wade+2011, Grunhut+2012a} is in the same order as the fraction of merger products that we expect. While this may be merely coincidence, several authors have speculated about the connection magnetic fields and binary interaction and mergers in particular \citep{Ferrario+2009}.

\subsection{Characteristics of binary products}

Apart from statistical statements about the frequency of binary product, we can consider ways to identify individual binary products. There are several characteristics that give hints for a binary origin.   Not all binary products are expected to show each of these characteristics and none of these characteristics uniquely signifies binary interaction.    

(a) The surface abundances of binary products may show signatures that indicate mixing or accretion of enricher gas.  In particular, a depletion of fragile elements is expected, because these can only survive in the cooler outer most layers of the star.  This concerns lithium, beryllium, boron, and fluorine \citep[e.g.][]{Langer+2010}.  In more sever cases the burning products of hydrogen fusion can appear at the surface, i.e. an enhancement helium, nitrogen and sodium and a depletion of carbon and oxygen. 

(b) Binary products are likely to have peculiar rotation rates. In principle, rapid rotation is expected as a result of spin-up during mass accretion or a merger \citep{Dufton+2011, de-Mink+2013-paper}, of which Be/X-ray binaries are direct evidence \citep[e.g.][]{Rappaport+1982}.   However, if binary products experience strong angular momentum loss, for example through magnetic braking, this may result in very slow rotators. 

(c) If binary interaction occurred recently there may be signs of shedded material in the circum-stellar medium,  either a (bipolar) ejection nebula  \citep[as in the case of the promising candidate for a merger product, the magnetic O6.5f?p  star HD 148937, ][]{Leitherer+1987, Smith+2004, Naze+2008, Naze+2010} or a circum-binary disk  \citep[as seen for example around the interacting system RY Scuti,][]{Grundstrom+2007}. The typical lifetime of such a nebula is expected to be around $10^4$\,yrs.  Therefore one can expect several stars that have merged recently and thus still display a circum-stellar remnant \citep{Langer2012}. 

(d) It has been suggested that a merger process may lead to the generation of a magnetic field \citep[e.g.][]{Tout+2008, Ferrario+2009}.  Strong, large-scale fields can be detected through spectropolarimetry, as for example for the companion of Plaskett star that has recently been spun up by mass accretion \citep[e.g.][]{Grunhut+2013}.  

(e) The supernova explosion of the primary star can break up the binary system, depending on the amount of mass lost from the system and the magnitude and orientation of the birth kick of the compact object \citep{Blaauw1961, Hoogerwerf+2001}.   When a tight system is disrupted the companion will acquire a high velocity. Such ``runaway stars'' can be detected by measuring the radial velocity, the proper motion or by indirect evidence such as the presence of a bow shock or its remote location away from regions of star formation.   In many cases the acquired velocities will be moderate, i.e. a few to tens of \kms\, \citep[e.g.][]{Eldridge+2011}. These stars would not be classified as runaway stars; ``walk-away star'' may be a more appropriate term.  However, they can travel tens to hundreds of parsecs from their birth location, because $1\kms \approx 1 \,{\rm pc}\,{\rm Myr}^{-1}$ and these binary products may live for tens of Myr before they explode. 

(f) Binary products are not expected to have a nearby unaffected main-sequence companion.   

(g)  If the former companion is still present it may be detectable in some cases through an UV excess \citep[e.g.][]{Gies+1998, Peters+2013}, in the case of a stripped helium star or through its X-ray properties when the former companion is an accreting compact object.

(h) Within coeval stellar populations such as star clusters, the binary products may stand out as the most luminous objects, possibly appearing younger than the age of the cluster. In this case they are the  massive analogue of blue stragglers  \citep[][]{van-Bever+1998, Chen+2009a}. For young star clusters, where the turn-off is not well defined, binary products may still stand out as the upper mass tail of the stellar mass function \citep{Schneider+2013}.

\section {Conclusion and Discussion} \label{Scon}

Based on our simulations, we predict that a population of early-type stars characterized by continuous star formation  is contaminated by stars that have experienced interaction with a companion.  We estimate the fraction of binary products in such a sample to be $30^{+10}_{-15}$ \% and the fraction of mergers to be $8^{+9}_{-4}$\%.  The error bars quoted here refer to the largest variation we obtain when varying the input distributions and the treatment of the physics of binary interaction with respect to our standard model. Even though larger variations cannot be excluded if multiple assumptions need to be adjusted in a way that systematically favors or disfavors binary products,  we conclude --- given our current understanding --- that the contamination of a sample of early-type stars with binary products is considerable. This poses a potential challenge when using these samples for various applications. . 

In particular, our findings raise questions about the validity of tests and calibration of single stellar models against observational samples.   We have shown that the commonly adopted strategy of excluding detected binaries from a sample is counter-productive.  Removing systems with detectable radial velocity variations preferentially removes binary systems that have not yet interacted from the sample. Post mass-transfer systems and mergers are left in the sample, accounting for $40^{+25}_{-20}$\% in total, mergers in particular account for $15^{+20}_{-10}$\%.

Our findings also shed new light on the interpretation of classes of peculiar stars by raising the question whether binary interaction or mergers are responsible for the peculiarity. While objects such as binary products and stellar mergers in particular may sound exotic, we predict that they are quite common.   This also raises concerns about our understanding of resolved and unresolved stellar populations, in particular the accuracy with which we can derive quantities such as the star formation rate, mass-to-light ratio and initial mass function using population synthesis models which do not account for the effects of binarity.

Our results emphasize the need further constrain the distribution of initial parameters, in particular the distribution of initial orbital periods. Concerning the physical assumptions the certainty of our predictions is mainly limited by our poor understanding of the mass transfer efficiency and the treatment of contact systems and mergers. Concerning the design of observational surveys our findings call for prioritizing efforts to devote great care to a careful and systematic inclusion or selection against known binary systems.   

\acknowledgments{{\it Acknowledgements:} \footnotesize 
We acknowledge various members of  the VLT-FLAMES massive stars consortium, PI: C. Evans for stimulating discussions.  SdM acknowledges support for this work by NASA through an Einstein Fellowship grant, PF3-140105 and a Hubble Fellowship grant HST-HF-51270.01-A awarded by STScI, operated by AURA, Inc., for NASA, under contract NAS 5- 26555.}

\bibliography{my_bib}

\end{document}